\documentclass[prl,twocolumn,,aps,showpacs]{revtex4} 

\usepackage{amsmath}
\usepackage{graphicx}
\usepackage{dcolumn}
\usepackage{bm}

\begin{document}


\title{A fast and efficient gene-network reconstruction method from multiple over-expression experiments}

\author{Dejan Stoki\'c$^{1}$, Rudolf Hanel$^{1}$, Stefan Thurner$^{1,2}$}

\affiliation{
$^1$ Complex Systems Research Group; HNO; Medical University of Vienna; 
W\"ahringer G\"urtel 18-20; A-1090; Austria \\
$^2$ Santa Fe Institute; 1399 Hyde Park Road; Santa Fe; NM 87501; USA\\
} 


\date{Version \today}

\begin{abstract}
Reverse engineering of gene regulatory networks presents one of the big challenges in systems biology. 
Gene regulatory networks are usually inferred from a set of single-gene over-expressions 
and/or knockout experiments. Functional relationships between genes are retrieved either from 
the steady state gene expressions or from respective time series. We present a novel algorithm for 
gene network reconstruction on the basis of  steady-state gene-chip data 
from over-expression experiments. The algorithm is based on a straight forward solution of 
a linear gene-dynamics equation, where experimental data is fed in as a first predictor 
for the solution. We compare the algorithm's performance with the NIR algorithm, 
both on the well known {\it E.Coli} experimental data and on in-silico experiments. We show superiority of the proposed algorithm in the number of correctly reconstructed 
links and  discuss computational time and robustness. 
The proposed algorithm is not limited by combinatorial explosion problems and can be used 
in principle for large networks of thousands of genes.
\end{abstract}

\maketitle
 
\section{Introduction}

Prediction of functional relationships between genes, starting from actual gene expression data, 
is one of the primary goals of systems biology. Despite large efforts in this direction 
\citep{gardner1,markowetz}, either based on transcription factor -- 
promoter interaction \citep{bussemaker,tavazoie}, or  on inferring gene networks 
\citep{chen,dhaeseleer,akutsu,yamanaka,gardner}, methods  for reliable predictions 
of collective behavior of gene-activity are yet to be found. 
Some general facts about the topology of gene regulatory networks \citep{sneppen, jeong2, jeong}, 
statistics of gene expressions \citep{zivkovic} or the dynamics of gene regulation \citep{chen} are becoming to be  
understood. This knowledge is far from sufficient to successfully reconstruct gene networks, 
but can be helpful in limiting the tremendous number of parameters involved in reconstruction. 
Even if the average degree of the gene regulatory network\footnote{The degree $k_i$ of a node $i$ in a network is defined as the 
number of links that emerge from -or point to- that node. 
The average degree is denoted by $\langle k \rangle$.},  i.e. the number of genes regulated by some gene on average, 
was known, noisy and limited data will always lead to severe problems. 

There are basically two types of reverse engineering approaches depending on the experimental 
setup, inferring the gene network from steady-state \citep{gardner,yamanaka} or from 
time-series \citep{yeung,arkin} experiments. By using steady-state experiments, 
one can not draw any conclusion about the dynamics of gene regulation. 
Conducting time-series experiments gives helpful insights into gene regulatory dynamics, 
but often with the price of getting redundant information. 
Further, due to costs full time-series data on gene expression are in general not available. 
As described in \citep{gardner1}, one can further divide the reverse engineering methods 
into four categories: differential equation models \citep{chen, dhaeseleer}, boolean network models \citep{akutsu}, Bayesian network models \citep{yamanaka} and association networks \citep{fuente}.

How can gene regulatory network reconstruction methods be validated and compared? 
Neither a standardized biological benchmark, nor a consensus on what class of models to use for 
in-silico testing exists \citep{gardner1}. 
The usual way is to validate a method either by applying it on a given experimental dataset 
or on in-silico datasets. In both cases one has to deal with different problems. Applying a method to an 
experimental dataset, poses the problem of comparing the reconstruction result with a network which is 
always just a consensus on how a biological network {\it could} look like, 
but never the exact gene regulatory network. 
On the other hand, when applying a reconstruction models to in-silico data, one has a perfect reference network, 
however the generated timeseries data is a result of a dynamical model of gene interaction,  
which cannot be shown to overlap with the real gene regulation dynamics. 
We decided to selected an {\it E.Coli} dataset \citep{gardner} as a satisfactory biological validation of our 
algorithm, because the underlying SOS response network has been subject to over 30 years of 
research, which provides us with a reasonable consensus about the actual gene regulations going on in that 
particular sub-network. For in-silico validation, we used a gene regulation model proposed in \citep{stokic}. 
This model simultaneously captures a series of experimental facts, such as the distribution of genome wide 
gene-expression levels , multi-stability and periodicity.

In the following we introduce a novel reverse engineering algorithm and compare it with 
\textit{Network identification by multiple Regression} (NIR) \citep{gardner}.  NIR has been 
applied to  the same SOS response network of {\it E.Coli}, which makes it possible to compare 
the two algorithms. Further NIR is considered as the state of the art algorithm which 
has so-far  not been outperformed in the quality of reconstruction. There are faster algorithms than NIR, 
which suffers from a binomial explosion problem, and is thus limited to relatively small 
networks. One fast algorithm was presented by \citep{mogno}, another recent algorithm 
\citep{yamanaka} claims the ability to reconstruct 
links with better statistical significance. 

The idea of the NIR algorithm is to reconstruct the network by using a 
least-squares regression approach, where RNA concentrations are regressors 
and the external perturbation  is the dependent variable. 
NIR enforces the same number of regulatory links to every gene, which is clearly unrealistic.  
The ensemble of links which provides the least squared error is selected as the 
optimal solution. For a  given average degree $\langle k\rangle$,  
the least-squares error is calculated for all the $ N\choose\langle k\rangle$ 
combinations, where $N$ is the size of the network. This becomes a combinatorial problem for 
large networks.

\section{Reconstruction method}

A system of interacting genes can be seen as a complex network, 
where every directed link represents a functional relationship between 
two genes. For simplicity, let us assume that this link will contain both 
transcriptional and translation levels of gene interactions. 
In this oversimplified view one can assume that the gene expression level changes in time as
\begin{equation}
 \dfrac{dX}{dt} = g(X(t))\quad,
 \label{unknownf}
\end{equation} 
where $g(X)$ is an a priori unknown function of a time dependent vector of gene 
expression levels $X$. If there are $N$ genes in the (sub)network under study 
(e.g. $N$ genes on a custom chip), vector $X$ 
has $N$ components. If we assume, as in \citep{yeung}, that $g(X(t))$ is a linear 
function (or after linearization of a more complicated function) one can write Equation (\ref{unknownf}) as
\begin{equation}
 \dfrac{dX}{dt} = AX + \mu\quad,
 \label{linearf}
\end{equation} 
where $\mu$ is a vector of gene over-expressions and $A$ is a constant adjacency matrix, containing the "strength" of gene-gene interaction. The elements $A_{ij}$ can be positive or negative real numbers, indicating activating or inhibiting interactions,  respectively.
%
%
By solving Equation (\ref{linearf}) one formally gets
\begin{equation}
  X^{\mu}(t) = e^{At}X^0 + A^{-1}(e^{At}-I)\mu\quad,
 \label{linsolution}
\end{equation} 
Where superscript $\mu$ indicates the system was perturbed with the constant vector $\mu$. After $M$ over-expressions, one can write the above equation in matrix form
\begin{equation}
  \hat{X} = e^{At}\hat{X}_0 + A^{-1}(e^{At}-I)\hat{\mu}\quad,
  \label{linsolutionM}
\end{equation} 
where $\hat{X}$ is the collection of all gene expression levels after the $M$ over-expressions experiments, 
organized in a $N\times M$ matrix, where one of the 
$M$ columns is a time dependent $N$-vector of gene expression levels for different gene being
over-expressed. In the following let us assume that we are able to perform 
$N$ over-expression experiments, i.e. $M=N$.  
$\hat{\mu}$ is a $N\times N$ diagonal matrix of gene over-expression levels. $\hat{\mu}$ is diagonal, because in every over-expression experiment just one gene is being over-expressed 
(which is the experimentally feasible case). At this point we emphasize that even though we know from the 
way over-expression experiments are prepared that the matrix 
$\hat{\mu}$ is diagonal, one often has little to no experimental control about the exact amplitude
of its entries. This problem is mitigated for small times $t\ll1$. To see this we define 
\begin{equation} 
Q\equiv\frac{1}{t}(\hat{X}(t)-e^{At}\hat{X}_0)\hat{\mu}^{-1}\quad.
\label{theQmat}
\end{equation}
Using this definition and abbreviating $\bar A\equiv At$ Equation (\ref{linsolutionM}) can be rewritten as
\begin{equation} 
\bar A = \ln\left( I+\bar AQ\right).
\label{implicitA}
\end{equation}
It is easy to check that in the short time limit
\begin{equation}
\lim_{t\rightarrow 0} Q = I
\end{equation}
holds. For very short times $t$ our lack of knowledge
is thus  basically irrelevant and estimating $\bar A Q$ reduces to estimating $\bar A$.
It is also not hard to realize that for sparse adjacency matrices $\bar A$ the
relative response
\begin{equation}
  Y_{ij} \equiv \dfrac{X_i^j-X^0_{i}}{X^0_{i}},
\end{equation}
will provide a good first estimate, i.e. $\bar A\propto Y$,
where $X^0_i$ is the gene expression level of the $i$th gene when 
no perturbation has occurred ($\mu\equiv 0$) and $X_i^j$ the gene expression level of the $i$th 
gene, where the $j$th gene has been over-expressed. 
Moreover, linearity assures that relative responses for short times will be  small, $|Y_{ij}| \ll1$.

However, for times in the order of a cell-cycle $t\sim 1$ and less sparse matrices 
these estimates will not be sufficient. 
The idea is to replace $Y_{ij}$ by some function $D_{ij}\equiv f(Y_{ij})$ which has the properties that (i)
$f(Y_{ij})\sim Y_{ij}$ for $|Y_{ij}| \ll 1$ and (ii) $f$ is a monotonously increasing function.
Since in practice $Y_{ij}$ can range over many decades in amplitude we also presume that (iii) $f$ should be a concave
function. Lastly, (iv) $f$ has to be defined on $[-1, \infty]$ since $-1\leq Y_{ij}$, but in principle could be arbitrary
large for positive values. Maybe the simplest function fulfilling this requirements is the logarithm, thus
we can estimate $\bar A Q = D$ for $f(x)\equiv \ln(1+x)$, i.e
\begin{equation}
D_{ij}=\ln\left(\dfrac{X_i^j}{X^0_{i}}\right).
\end{equation}
This means that we effectively estimate $\bar A=\ln(I+D)$, where $I$ is 
the identity matrix. For the matrix logarithm to provide
unique solutions, $I+D$ should not have any negative real 
eigenvalues. Since experimental results show that this is not the case in 
general we use a {\em cleaned} version (see below) of $D$, denoted by 
$D^0$ such that $I+D^0$ has no negative real eigenvalues and the prediction 
of the adjacency matrix is given by
\begin{equation}
 \bar A = \ln\left(I+D^0\right)\quad.
 \label{assumption}
\end{equation}

\subsection{Eigenvalue cleaning}

In general, the logarithm of a matrix can have an infinite number of real and complex solutions. 
In order to find a unique solution of $\ln(D+I)$, matrix $D+I$ can not have negative real eigenvalues. 
If we take a look at the eigenspectrum of matrix $D$ from various experiments, both 
biological and in-silico, we notice that most of the eigenvalues are complex, however a 
small number of eigenvalues are real, both positive and negative. Thus, we first have to 
\textit{clean} the matrix $D+I$, meaning to set all the negative eigenvalues to small 
positive number $\epsilon$. This is done by first diagonalizing matrix $D$:
\begin{equation}
\tilde D= U^{-1}DU=diag(d_1,...,d_N) \quad, 
\end{equation}
where the eigenvalues are ordered in a way that first $L$ eigenvalues are real and less then -1, $d_{i}^*= d_{i}<-1, \forall i\leq L$.
These $L$ diagonal elements are set to $\epsilon-1$
\begin{equation}
\tilde{D^{\epsilon}}+I = \left(
  \begin{array}{c c c c c c}
     \epsilon & 0 & \ldots & 0 & \ldots & 0\\
     0 & \epsilon & \ldots & 0 & \ldots & 0\\
	 \vdots & \vdots  & \ddots & \vdots &  & \vdots\\
     0 & 0 & \ldots & d_{L+1} & \ldots & 0\\
     \vdots & \vdots  &  & \vdots & \ddots & \vdots\\
     0 & 0 & \ldots & 0 & \ldots & d_{N}\\
  \end{array} \right),  
\end{equation}  
and are rotated to yield the cleaned matrix
\begin{equation}
  D^0 = U\tilde D^{\epsilon}U^{-1}.
\end{equation} 
Matrix $D^0+I$ no longer has negative real eigenvalues, and a unique prediction of an 
adjacency matrix $A$ - reconstructed gene regulatory network - can be given
\begin{equation}
A = \ln(D^0+I).
\end{equation}

\subsection{Thresholding}

Our solution $A$ will in general represent a fully connected network, with a certain 
distribution of link weights around zero. The reason why we are always getting fully 
connected network, e.g. network without zero entries in adjacency matrix, is because of 
the noisy measurements. Real gene regulatory networks are never fully connected, 
but are characterized by an average degree $\langle k\rangle$, which has been estimated to 
be relatively small $\sim 2-4$ \cite{jeong}. 
For simplicity we assume $\langle k\rangle$ for the undirected unweighted case. 
Knowledge of $\langle k\rangle$ allows to define a clear thresholding scheme. 
All entries in $A$ below a threshold $\alpha$ are set to zero. $\alpha$ is chosen such, that
matrix $A^0_{ij}$ has the average degree $\langle k\rangle$, i.e. 
\begin{equation}
  A^0_{ij}\equiv A_{ij}\theta(\vert A_{ij}\vert-\alpha),
  \label{impose0}
\end{equation} 
such that
\begin{equation}
  \frac{1}{N}\sum_i\sum_j\theta(\vert A^0_{ij}\vert) = \langle k\rangle,
\end{equation} 
$A^0$ is the first approximation of the gene regulatory network we want to reconstruct.

\subsection{A note on fewer than $N$ experiments}

In the case where the number of over-expression experiments $M$ is lower than the 
size of the network $N$, matrix $D$ is not quadratic, thus we are unable to 
calculate the matrix logarithm. Information about the influence of  
gene $j$ ($j>M$) on gene $i$ is missing. A way around is that one 
can introduce a measure of the distance between two  genes in the network. 
Although the correlation between gene expressions in different over-expression 
measurements can not lead to any conclusion about the functional relations among 
genes, it can provide a good measure for the distance between the genes in the network. 
One can therefore simply calculate a matrix of correlation coefficients 
and replace the missing terms in $D$:
\begin{equation}
D_{ij} = \dfrac{N\sum_k D_{ik}D_{jk}-\sum_k D_{ik} \sum_k D_{jk}}{\sqrt{N\sum_k D_{ik}^2-(\sum_k D_{ik})^2}\sqrt{N\sum_k D_{jk}^2-(\sum_k D_{jk})^2}}
\end{equation}
Here the first index in $D_{ij}$, $i$ runs in the domain $M<i\leq N$, the second index, $1\leq j\leq N$.

\section{Testing the method}

We compare our results with the NIR algorithm both on an \textit{in-silico} dataset, 
as well as on the \textit{E.Coli} SOS response network \citep{courcellec,walker,koch,henestrosa,karp}, 
in the same way as in \citep{gardner}. We measure performance in two ways, firstly, 
by counting the fraction of correctly reconstructed positive, negative and zero links, 
denoted by $F_+$, $F_-$ and $F_0$, respectively. For later use we define 
$F \equiv F_++F_-+F_0$. Secondly, by calculating the extended Matthews correlation 
coefficient \citep{gorodkin}, a discrete version of Pearson's correlation coefficient, 
extrapolated onto  $K\times K$ confusion matrices. Matthews  
correlation coefficient is taking values in the interval $[-1,1]$, where $0$ stands for 
no correlation between predicted and real case, and $1$ and $-1$ stands for 
complete or negative correlation respectively. The $K$-category correlation coefficient is defined as
\begin{equation}
R^K=\tfrac{\sum_{klm}(C_{kk}C_{lm}-C_{kl}C_{mk})}{\sqrt{\sum_k(\sum_l C_{kl})(\sum_{l'\\k'\neq k}C_{k'l'})}\sqrt{\sum_k(\sum_l C_{lk})(\sum_{l'\\k'\neq k}C_{l'k'})}},
\label{RK}
\end{equation}
where $C$ is a $K\times K$ confusion matrix, or more precisely the element $C_{kl}$ is counting the number of cases where category $k$ is predicted, but category $l$ was present. In our case, $K=3$ and the categories are: positive link, negative link and no link between any two genes. 
It is straight forward to see that p-values for any value of $R^K$ can be computed exactly in the same way as for the 
Pearson correlation coefficient, provided sample size is given. 

It is important to stress the difference in measuring reconstruction performance in \textit{in-silico} 
and biological experiments. While in biological networks, self-regulation is a part of the 
complete gene regulatory network, in numerically simulated gene regulation dynamics, 
self-regulation is often screened by negative self-degradation rates, which have to be imposed, 
in order to keep the dynamics sufficiently stable, see e.g. \citep{stokic}. 
To be as correct and conservative as possible, we therefore compare our reconstructed adjacency matrix 
{\em only} with the off-diagonal 
elements in the \textit{in-silico} case. In the \textit{E.Coli} case we of course 
compare with the complete adjacency matrix.

\subsection{In-silico testing}

\begin{figure*}[h]
\resizebox{1.0\textwidth}{!}{\includegraphics{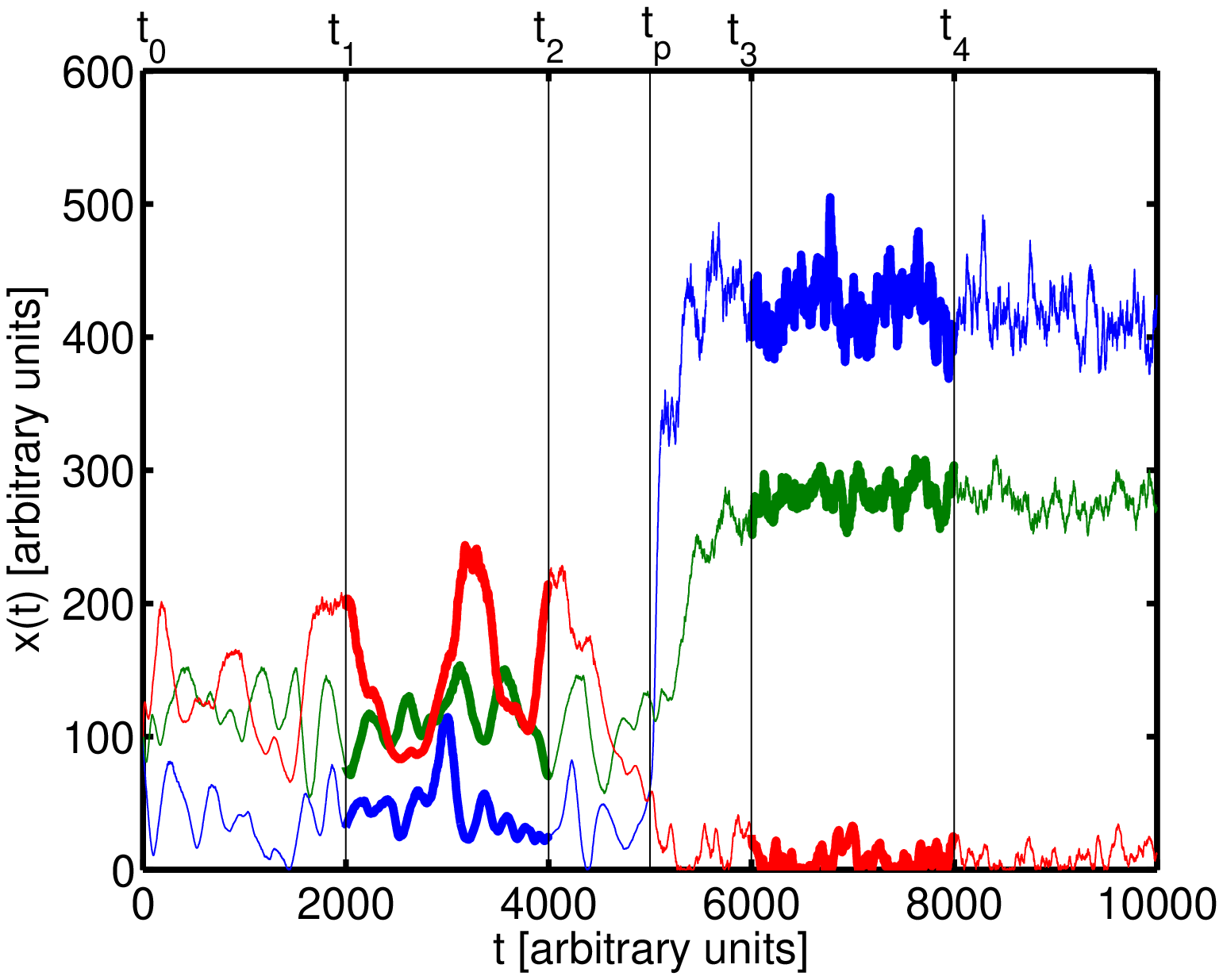}}
\caption{Time series of three randomly selected trajectories (numerical solutions
of Eq. (\ref{model})), showing the measurements of gene expression levels in \textit{in-silico} over-expression experiment. Gene expression levels were measured from time $t_1$ until $t_2$ for the steady state levels, and from time $t_3$ until $t_4$ for the effect of perturbation. At the time $t_p$ one gene was over-expressed.}
\label{tseries}
\end{figure*}

We employ a recently proposed dynamical gene-gene interaction model, which is able to capture a 
series of experimental facts on gene-expression statistics \citep{stokic}: (i) distribution of 
gene-expression increments over time, (ii) multiple equilibria, (iii) stability. The model is defined as
\begin{equation}
 \frac{d}{dt} x_i(t) = \sum_j A^{model}_{ij}( x_j(t) - x^0_j) + \xi_{i}(t) ( x_j(t) - x^0_j) + \eta_i \quad ,   
 \label{model}
\end{equation}
with a positivity condition imposed for gene expression levels (non negativity of concentrations):
\begin{equation}
x_{i}(t) \geq 0  \quad \forall i \quad.
\label{poshalfsp}
\end{equation}
Here, $A^{model}$ is a real valued adjacency matrix of gene-gene interactions. 
It is modeled as a particular random matrix, mimicking experimentally known facts \citep{stokic}. 
$x(t)$ is a vector of gene-expression levels in time $t$, constant vector $x^0$ indicates steady 
state gene-expression levels. $\xi$ and $\eta$ are multiplicative and additive noise terms, respectively, 
which are a generic feature in chemical reactions. 
Using the dynamics defined in Equation (\ref{model}) we generate the time series of 
gene expression levels $x(t)$, and simulate the effects of perturbation by adding a 
constant perturbation vector to the Equation (\ref{model}). For details, see \citep{stokic}. 
We \textit{measure} the gene expression levels as time averages over concentrations: 
$X^0_{i}=\dfrac{1}{t_2-t_1}\int_{t_1}^{t_2}x_i(t)dt$ and 
$X_i^j=\dfrac{1}{t_4-t_3}\int_{t_3}^{t_4}x_i(t)dt$,  where 
$t_0<t_1<t_2<t_p<t_3<t_4$.  $t_0$ is the initial time point of the simulation (after discounting transient behavior), 
$t_p$ is the time at which the perturbation vector (with the $j$th component being non zero) 
is applied. The procedure is depicted in Figure \ref{tseries}.

\subsection{Testing on the \textit{E. Coli} dataset}

We use the wild-type {\it E.Coli} strain MG1655 available at \citep{gardner}. 
The reason for testing our method on this particular dataset is the fact that the SOS 
response of the {\it E.Coli} is well understood, and some consensus over the topology of its gene regulatory network is reached. Moreover it is possible to compare reconstruction success with other groups \citep{mogno,yamanaka,supper}.
We test the performance by counting the fraction of the correctly reconstructed 
links of all three classes (positive, negative and zero), 
and with the extended Matthews correlation coefficient.

\subsection{The pure-chance reconstruction threshold}

A strong criterion of checking the performance of any reconstruction method we consider, is to compare 
it with a pure random-reconstruction. Several proposed gene network reconstruction algorithms 
can be shown to perform only slightly above pure-chance reconstruction. 
Random reconstruction can be performed in the following way.
Suppose that $\langle k\rangle$ denotes the true average degree  of 
the network, which may or may 
not be known, and $k_g$ denotes a guess on  $\langle k\rangle$. 
Since we estimate that the directed network has $L=Nk_g$ links we take a 
fully connected network and assign a random order to all $N(N-1)$ links. 
Then we take a random number with three outcomes: $+$ (positive weight), 
$-$ (negative weight), and  $0$ (no link), and assume that there are as many positive as negative 
links. The distribution of these outcomes therefore is such 
that both $+$ and $-$ occur with probability $w_\pm=k_g/2N$, while the $0$ appears with 
probability $w_0=1-k_g/N$. 
The {\em true} probabilities, i.e. the probability of $+,-,0$ if the true average $\langle k\rangle$ was known, 
however are, $p_\pm=\langle k\rangle/2N$ and  $p_0=1-\langle k\rangle/N$.
Now we pick one link after another in the given random order,  and assign a random  symbol, $+,-$ or $0$
and repeat this until $L$ links have been assigned either $+$ or $-$.
Since 'throwing the dice' is an event independent of the network topology, 
one can simply compute $F_\pm^{rand}(k_g |\langle k\rangle)=w_\pm p_\pm$ and
$F_0^{rand}(k|\langle k\rangle)=w_0 p_0$.

If reconstruction is  based on pure chance 
the expected $K$-category correlation will be $R^K=0$.
This can be seen by inserting the confusion matrix $C_{ij}=w_i p_j$, $i$ and $j$ indexing $+,-$ or $0$, 
into Equation (\ref{RK}).

\section{Results}

\subsection{Reconstruction on in-silico data}

\begin{figure*}[h]
\resizebox{1.0\textwidth}{!}{\includegraphics{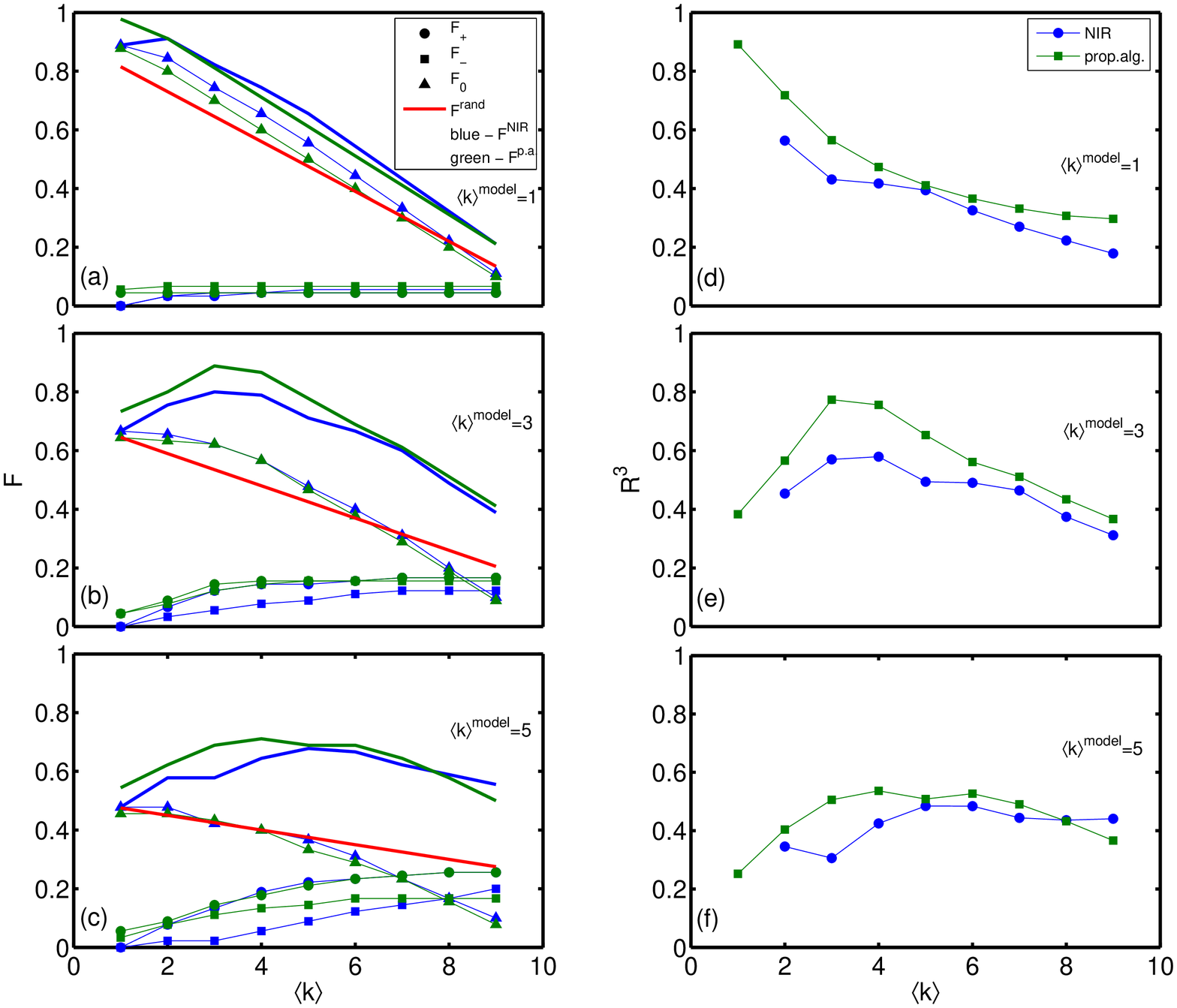}}
\caption{Results of network reconstruction for 
the proposed  algorithm (green lines) and NIR (blue lines) for in-silico experiments. The results 
both for the fraction of correctly reconstructed links ((a)-(c)) and extended Matthews correlation 
coefficient ((d)-(f)) are shown. Three in-silico networks with different average degree were 
constructed, for $\langle\tilde k\rangle$ equals 1((a),(b)), 3((b),(e)) and 5((c),(f)). 
In the plots where the fraction of correctly reconstructed links are shown, circles 
denote the fraction of positive links $F_+$, squares the fraction of negative links 
$F_-$ and triangles no links $F_0$. The red line represents the gambling threshold $F^{rand}$.}
\label{insilico}
\end{figure*}

We generated three different networks ($N=10$) with different 
connectivities ($\langle\tilde k\rangle \in \{1,3,5\}$), for purposes of in-silico testing of our 
reconstruction algorithm. Using a fixed adjacency matrices $A^{model}$ of these 
networks, we simulated time series of gene expression levels (see Figure \ref{tseries}) 
according to Equation (\ref{model}), with noise levels $\sigma = \bar\sigma=0.1$, where $\xi_i\in N(0,\bar \sigma)$ 
and $\eta_i\in N(0, \sigma)$. For details \citep{stokic}. As described in previous section, 
we measured the steady state gene expression levels before and after the perturbation of 
each gene in the network, denoted by $X^0_i$ and $X^j_i$, respectively. The so generated 
data was taken as an input for both reconstruction methods. In this case the exact value 
of the over-expression vector $\mu$ was used as an extra input parameter for the NIR reconstruction. 
In reality this exact value remains unknown. Results were produced for 20 
statistically identical realizations of networks for every connectivity $\langle\tilde k\rangle \in \{1,3,5\}$. 
All the networks provided very similar results, only one for every connectivity is shown in Figure \ref{insilico}. 
Here we compare the results of our reconstruction method with the  NIR algorithm for in-silico 
experiments. The left panel of the figure shows the fraction of correctly reconstructed links, 
for every link type ($F_+$, $F_-$ and $F_0$) as well as their sum $F$. 
The colors blue and green represent the NIR and the proposed method, respectively. 
The pure-chance threshold is shown to emphasize the significance of the result. 
The right panel shows the extended Matthews correlation coefficient. 
For the Matthews correlation coefficient the pure-chance threshold is constant at zero.
It is clearly seen that for the fraction of correctly reconstructed links
our method performs about equally well than the NIR for very sparse 
networks ($\langle k \rangle^{model}=1$) and  outperforms it in 
in more densely connected networks. There, 
when looking at the fractions of correctly reconstructed links one notices 
a slightly better performance of our algorithm, while for the extended Matthews 
correlation coefficient the difference is much more notable. To understand this difference, 
one has to take a closer look at the type I and type II errors of both methods. 
While the NIR algorithm makes almost the same number of reconstruction 
errors of all types, there is a clear distinction in errors made by our reconstruction algorithm. 
The vast majority of errors are made by assuming that there is a link 
(positive or negative) between two genes, while in the real case there is none, 
and vice versa. Only a few mistakes are made where the real positive link is 
reconstructed as negative, or vice versa. This is an additional asset of the proposed  
reconstruction algorithm.

\subsection{Reconstruction of the E. Coli SOS network}

\begin{figure*}[h]
\resizebox{1.0\textwidth}{!}{\includegraphics{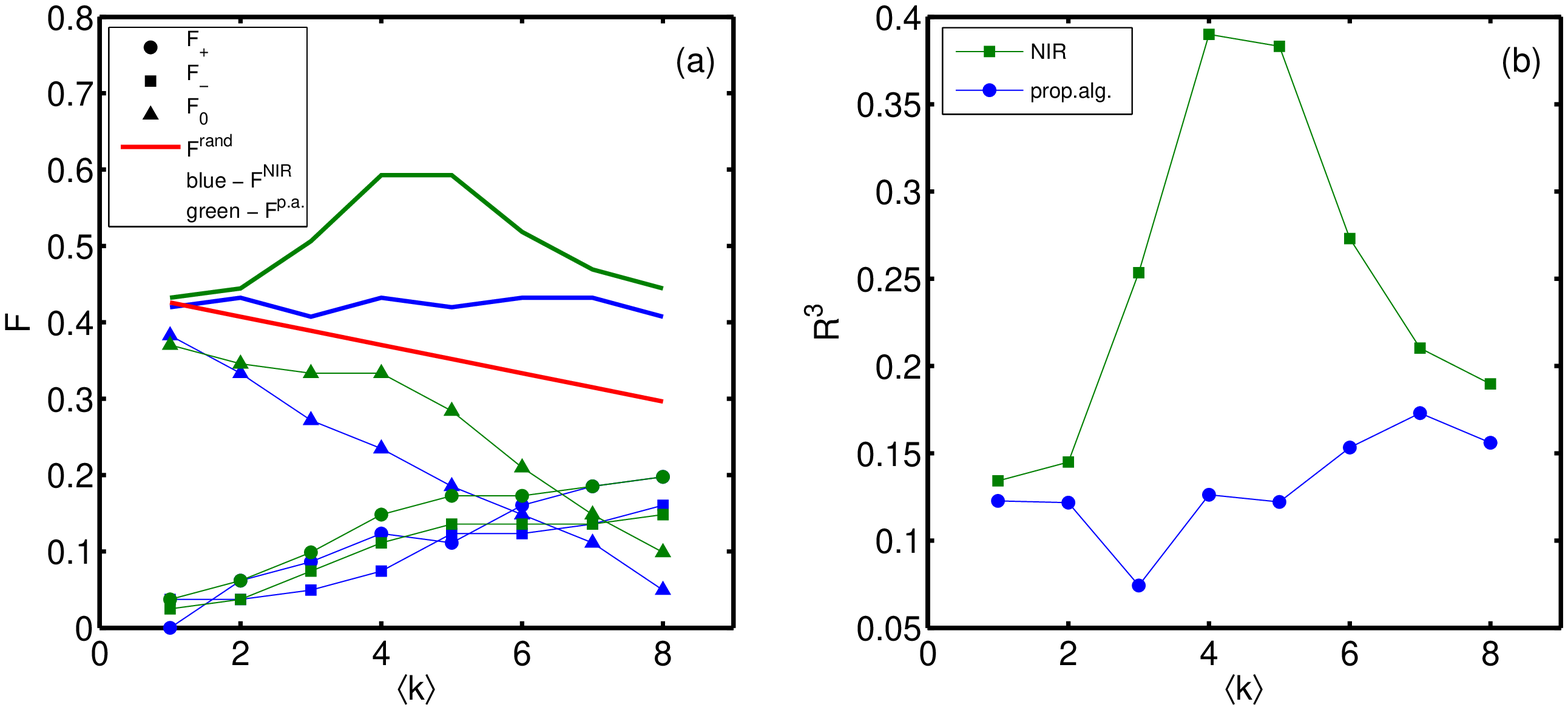}}
\caption{Reconstruction results  for the proposed algorithm (green lines) and NIR (blue lines) 
for E.Coli. The results both for the fraction of correctly reconstructed links (a) and extended Matthews 
correlation coefficient (b) are shown. In (a) circles denote the fraction of positive links $F_+$, squares 
the fraction of negative links $F_-$ and triangles no links $F_0$. The red line represents 
the gambling threshold $F^{rand}$.}
\label{ecoli}
\end{figure*}

Although our reconstruction method showed better results tested on in-silico networks than NIR, 
the true value of any reconstruction potential can be shown just on the real biological data. 
When testing both methods on {\it E.Coli} data, as shown in Figure \ref{ecoli} , our 
reconstruction method outperforms the NIR more visibly, in both performance measures. 
To stress the difference in the quality of reconstruction we present p-values 
of given correlation coefficients between the real and reconstructed networks are. 
Given the sample size $K=81$, i.e. the number of links to be reconstructed, 
and a $\langle k\rangle=4$ (known experimental value), the p-value of correlation coefficient $R^3_{NIR}=0.14$ for NIR is  
$p_{NIR}=0.2126$, while the p-value of correlation coefficient $R^3=0.4$ for our method is 
$p=0.002$. For $R^3$ values see Figure (\ref{ecoli}), at $\langle k\rangle=4$.  
Our reconstruction leads to a network which significantly correlates better with 
the experimentally known biological network.

One can easily notice that both reconstruction methods applied on in-silico data have 
their maxima in performance when the input average degree equals to the true one, 
$\langle k\rangle = \langle k\rangle^{model}$, which can be seen as an additional consistency check of the algorithm. 
On the other hand, after applying both reconstruction methods on  {\it E.Coli} data, just the proposed  
reconstruction algorithm shows its performance maximum at the $\langle k\rangle = \langle k\rangle^{E.coli}$ point, 
while the NIR method shows similarities in behavior to the pure-chance reconstruction.

The computational time needed to perform the NIR algorithm on this particular 9 node network is 
of order of magnitude of 1 minute, while our approach takes less than a second, both performed on 
a standard personal computer. The NIR algorithm is unable to cope with reconstruction of significantly 
larger gene regulatory networks, both from the time or memory consumption, while our method 
can deal with network sizes of up to realistic genomes.

Because of typically high levels of noise and uncertainty in biological data collected 
throughout actual experiments, the robustness of a method is of crucial importance. 
We tested both the NIR and our algorithm in the following way: in the in-silico experiments we generated 
a data matrix $X_1$ with $N$ genes and $N$ experiments. In this matrix we replace 
2 randomly chosen columns by random data (iid Gaussian entries with unit variance). 
This matrix we call $X_2$ and reconstruct networks $A_1$ and $A_2$ from $X_1$ and 
$X_2$, respectively. By comparing these two reconstructed networks, in the case 
for NIR reconstruction we find strong tendency of all links to change their position. 
In the proposed method links preferably change at the positions of the replaced columns.

\section{Discussion}

We introduced a reverse engineering procedure for gene regulatory networks, 
applicable on an experimental setup where all the genes belonging to a genetic (sub)network 
are being over-expressed one after the other, after which gene-chip measurements in the 
steady state are taken. We showed the reconstruction performance on both {\it in-silico} and 
biological data. The method is applicable to large networks, both from the computational 
memory or computational time point of view, which might be a problem for algorithms limited by 
combinatorial explosions. 

Except from technical benefits, the philosophy of our reconstructing method 
complies perfectly with the biological goals of conducting over-expression experiments. 
In contrary to the NIR algorithm or similar reconstruction methods, where the final 
solution is a network, where every link has same significance, our method ranks the reconstructed 
links by their influence,  which might be a very important issue in experimental 
gene interaction-detection 
Instead of randomly picking the links out of a given reconstructed  topology, here one 
can select interaction-links with the highest weights. This again ameliorates 
the consequences of not knowing the real network connectivity $\langle\tilde k\rangle$ a priori. 
While selecting a good value for $\langle k\rangle$ is crucial for getting reliable networks, 
it will not influence the ordering of the links by importance in the proposed algorithm. 
In other words, no matter which $\langle k\rangle$ is taken, the set of ranking of reliable  links 
will not change.

Another shortcoming of the NIR algorithm is the fact that the resulting network 
has a trivial, unrealistic degree distribution, a delta function, $\delta(k-k^{*})$. 
Thus, detecting genetic hubs, peripheral genes, or any other topologically important 
genes in the network is practically  impossible.  
The proposed method does not a priori restrict the topology of the reconstructed 
network except for the average degree $\langle k\rangle$ which is important for the thresholding 
only. 

The NIR algorithm needs as an input parameter for the successful reconstruction 
information external perturbation, which is in most of the cases 
just approximately known. In the in-silico experiments we have provided the 
exact information for NIR, however the NIR algorithm was still outperformed.

Supported by WWTF LS129 and Austrian Science Fund FWF Project P19132.

\end{document}